\documentclass[10pt,letterpaper,twocolumn]{article} 

\usepackage{ol2}
\usepackage[draft]{hyperref}
\usepackage{amsmath}

\begin{document}

\twocolumn[ 

\title{Crosstalk-free operation of multi-element SSPD array integrated with SFQ circuit in a 0.1 Watt GM cryocooler}


\author{Taro Yamashita,$^{*}$ Shigehito Miki, Hirotaka Terai, Kazumasa Makise, and Zhen Wang}
\address{
National Institute of Information and Communications Technology, 
588-2, Iwaoka, Nishi-ku, Kobe, Hyogo 651-2492, Japan 
\\
$^*$Corresponding author: taro@nict.go.jp
}
\begin{abstract}
We demonstrate the successful operation of a multi-element superconducting nanowire single-photon detector (SSPD) array integrated with a single-flux-quantum (SFQ) readout circuit in a compact 0.1 W Gifford-McMahon cryocooler. A time-resolved readout technique, where output signals from each element enter the SFQ readout circuit with finite time intervals, revealed crosstalk-free operation of the four-element SSPD array connected with the SFQ readout circuit. The timing jitter and the system detection efficiency were measured to be 50 ps and 11.4\%, respectively, which were comparable to the performance of practical single-pixel SSPD systems. 
\end{abstract}
\ocis{030.5260, 040.1240, 040.3060, 040.5160, 270.5570.}

] 

Superconducting nanowire single-photon detectors (SSPDs) are promising components in a wide range of applications, e.g., quantum key distribution, optical communications, and quantum optics, due to their many advantages \cite{Goltsman,Hadfield,Sasaki,Robinson,Natarajan}. Further improvement in performance is strongly desired, and one important feature is the counting rates limited by the large kinetic inductance $L_{k}$ resulting from the long nanowire. Although miniaturization of the active-area size effectively reduces $L_{k}$, the system detection efficiency (DE) simultaneously decreases due to inefficient optical coupling. In order to overcome this trade-off, a multi-element SSPD array is an attractive solution since this configuration can also provide spatial and pseudo-photon-number resolution \cite{Dauler}.

Readout technology is an important issue in developing a multi-element SSPD array system for practical use. The single-flux-quantum (SFQ) circuit is an adequate choice for readout electronics of the SSPD array because the required number of coaxial cables introduced from room temperature can be reduced in comparison to a conventional readout \cite{Terai1,Terai2,Miki}. By implementing SFQ circuits and SSPDs, the heat load from the room temperature through the coaxial cables is significantly reduced. Furthermore, the SFQ operates at high speed, low jitter, and low power consumption. Previously, we demonstrated the fundamental operation of single-pixel SSPDs integrated with a SFQ circuit in a cryocooler with a cooling power of 0.7 W \cite{Miki}. Thus, the next step should be a correct operation of a multi-element SSPD array integrated with the SFQ circuit in a compact closed-cycle cryocooler system from practical perspectives.

In the SSPD array system, undesirable interference between elements, or crosstalk, must not occur for proper operation. In \cite{Dauler}, the crosstalk of a two-element SSPD array was verified and no measurable crosstalk was observed in the conventional readout technique. On the other hand, in the SFQ readout once the output signals from each element are merged, it is unable to identify the reacted element, making crosstalk measurements difficult. To solve this problem, the output signals must be time-resolved after being merged in the SFQ circuit.

In this letter, we demonstrate the successful operation of a four-element SSPD array integrated with a SFQ readout circuit in a 0.1 W Gifford-McMahon (GM) cryocooler system, which is commonly employed for practical SSPD systems with single-pixel devices \cite{Miki15}. We verify the crosstalk and timing jitter by applying a new readout technique, which enables us to identify the signal for each element in a time-correlated single photon counting (TCSPC) measurement.

The SSPD array presented in this letter consists of a linear layout of four NbN meandering nanowires (elements A, B, C, and D in series) with a total area of 15 $\times$ 15 $\mu$m$^2$ \cite{Miki2}. The thickness of the film is 4.5 nm, and the nanowire width and period are 100 nm and 200 nm, respectively. An optical-cavity structure, as reported in \cite{Miki3}, was not attached to the device this time. The fabrication process is basically identical to that for the single-pixel SSPD, described elsewhere \cite{Miki4}. The superconducting critical temperatures are 10.4 K for all elements and the superconducting critical currents, $I_{c}$, are 38.1, 37.5, 39.1, and 39.9 ${\rm\mu}$A for the elements A, B, C, and D, respectively. We confirmed that there is no measureable crosstalk in $I_{c}$ between the elements \cite{Dauler}.

The SSPD array device was mounted on a fiber-coupled package, modified from the one for a single-pixel SSPD \cite{Miki2}. The package has four miniature RF connectors that can pass signals with a frequency up to 110 GHz. Each element in the package was connected to a corresponding connector through a coplanar waveguide line on the SSPD array chip. Incident light entered the device through a single-mode optical fiber from the back side of the chip, which also enables photon irradiation for optical-cavity SSPDs. In order to achieve efficient optical coupling, small gradient index lenses with a beam waist at the device of 8--10 $\rm\mu$m were fusion spliced to the tip of a fiber. Prior to cooling, optical alignment between the incident light spot and the active area was performed.

The SFQ readout circuit for the SSPD array was fabricated by SRL with a 2.5 kA/cm${^2}$ Nb standard process. In the circuit, there are four magnetically coupled DC/SFQ (MC-DC/SFQ) converters consisting of an input transformer connected to a 50 $\Omega$ load resistor and a DC/SFQ converter where SFQ pulses are generated at the rising edge of each input pulse. Details on the MC-DC/SFQ converter are described in \cite{Terai1}. 
Generated SFQ pulses transmit to a signal processing unit consisting of three confluence buffer (CB) gates, 
which has a function of multiplexing two input channels without clock pulses. 
As a result, the four input channels are multiplexed into a single channel. 
Merged SFQ pulses are sent to a voltage driver and converted into a rectangle pulse shape with an amplitude of 1.8 mV and a duration of 1.6 ns, easily detected using conventional electronics.

\begin{figure}[tb]
\centerline{\includegraphics[width=8cm]{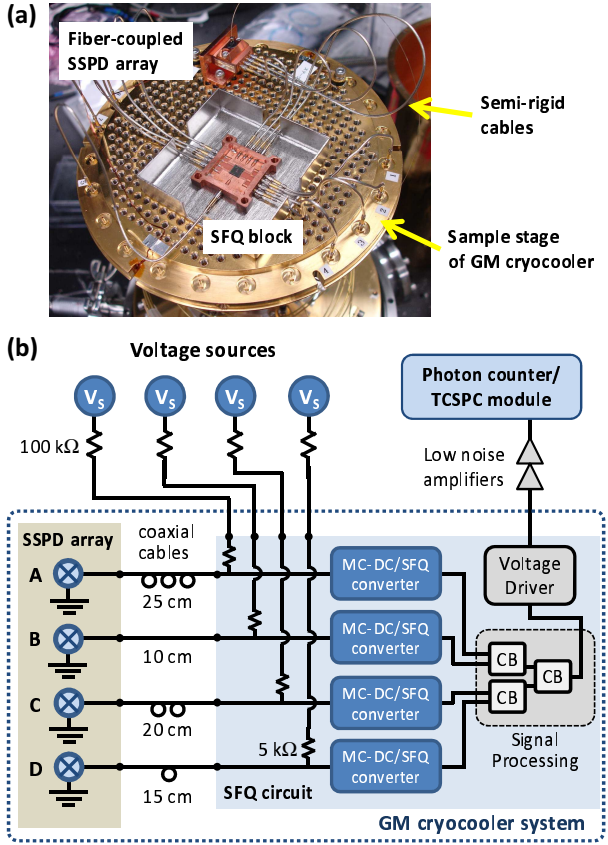}}
\caption{(a) Implementation of the GM cryocooler system. 
(b) Schematic setup of SSPD array with SFQ circuit. 
Each element and SFQ input are connected by coaxial cables of different lengths. 
For simplicity, four bias lines to drive SFQ circuit are not shown in the schematics.}
\label{fig:setup}
\end{figure}

The fiber-coupled SSPD array package and the SFQ circuit were implemented into a closed-cycle GM cryocooler system with a cooling power of 0.1 W at 4.2 K, as shown in Fig. \ref{fig:setup} (a). Nine brass semi-rigid coaxial cables were introduced into a cryocooler system: one cable for the output signal and eight cables for the bias current supplied to the SSPDs and the SFQ readout circuit. The eight coaxial cables for the bias can be replaced by dc cables with a much lower thermal conductivity in the future, reducing the heat load to the cryocooler system. The SFQ circuit was surrounded by a double $\mu$-metal shield in order to avoid vortex traps. The sample stage was cooled down to 2.35 K with a thermal fluctuation range of 5 mK. By applying a bias current of around 40 mA to the SFQ readout circuit, the sample stage temperature increased to 2.42 K, which is still low enough for the operation of both the SSPD and the SFQ circuits.

We first verified the crosstalk in the output signal counts of the each element. In order to identify the reacted element even after the output signals are merged in the SFQ circuit, we adopted a new configuration shown in Fig. \ref{fig:setup} (b). Each element of the SSPD array was biased independently by voltage sources, and connected to SFQ input ports using coaxial cables with different lengths: 10, 15, 20, and 25 cm for elements B, D, C, and A, respectively. Since the output pulses generated at the each element travel to the SFQ circuit through the coaxial cables with different length, the arrival times at the SFQ input port are different for each element. Assuming a delay time in the coaxial cable of 5 ns/m, a 5 cm difference in cable length gives an arrival time difference of ~250 ps, which is adequate to separate signals from other elements because the timing jitter of the SFQ pulse is expected to be around several tens of picoseconds \cite{Miki}.

\begin{figure}[tb]
\centerline{\includegraphics[width=8cm]{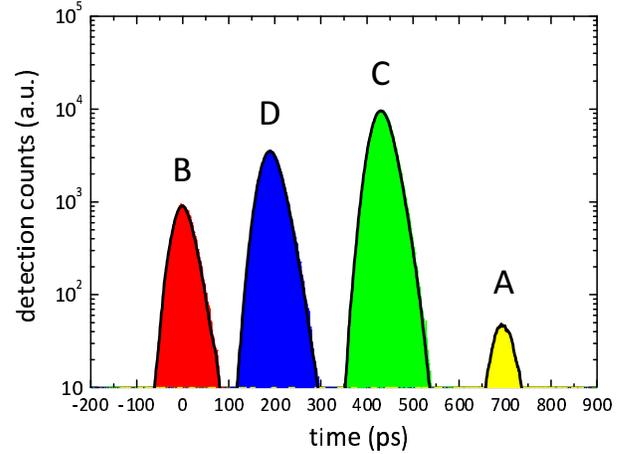}}
\caption{Histogram of time-correlated counts of SFQ output at a dark count rate of 100 cps. 
Colored bar graphs indicate the histogram of the SFQ output when one element was biased. Black lines 
indicate the histogram for all elements biased.}
\label{fig:jitter}
\end{figure}

Using this readout configuration, we measured the histogram of the time-correlated counts. In this measurement, we used a 1550 nm wavelength pulsed laser with 100 fs pulse width. The light power was heavily attenuated at around 0.1 photon per pulse at the optical input port of the cryocooler, so as not to react several elements at the same time. The time correlation between a synchronized trigger pulse and the output pulse from the SFQ circuit was measured by the TCSPC module with a resolution of 1 ps. Figure \ref{fig:jitter} shows histograms of the time-correlated counts from the SFQ output at the dark count rate of 100 cps. Colored bar graphs indicate histograms when only one element was biased. 
Output peaks with a full width at half maximum (FWHM) of 50--54 ps from each element appeared at different timing positions, corresponding to arrival time delays. Although there is a variation in peak levels for each element, this comes from a difference in the DE of the each element, as discussed later. The black line in Fig. \ref{fig:jitter} shows a histogram when all four elements were biased simultaneously. The output peaks from the four elements were clearly separated even when all elements were operated and merged by the SFQ circuit. Comparing the center position and the FWHM of the each peak to those biased individually, we obtained the difference of the center position within 2 ps and that of the FWHM within 2\%, both of which are smaller than the measurement noise. This means that all elements were unaffected by a bias current to the other elements. These results indicate that the SFQ circuit correctly merged the signals from all elements and produced output signals without any crosstalk.

\begin{figure}[tb]
\centerline{\includegraphics[width=8cm]{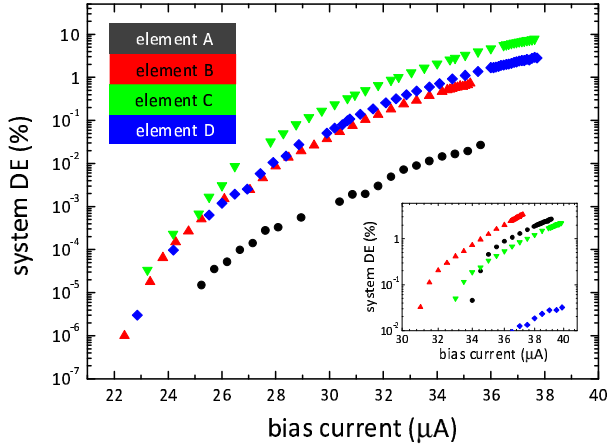}}
\caption{System detection efficiency versus bias current for the each element using the SFQ readout. 
Inset: system detection efficiency versus bias current 
after realigning light-spot position around element B.}
\label{fig:DE}
\end{figure}

Figure \ref{fig:DE} shows the bias current dependence of the system DE with the SFQ circuit. In this measurement, a 1550 nm wavelength continuous-wave laser diode was used as an input photon source. The output power was heavily attenuated to a level of 10$^6$ photons per second. The element C has the highest DE of 7.9\%, and the elements B and D, which are both sides of the element C, have a DE of 0.7\% and 2.8\%, respectively. On the other hand, element A has an extremely low DE of less than 0.1\%. The DE variation qualitatively agrees with the peak levels from each element in Fig. 2. This result indicates that the incident light spot shifted slightly from the center of the active area, and the spot center was positioned around element C. Since the diameter of the light beam waist is 8--10 $\mu$m, incident photons rarely entered element A. To confirm this, we realigned the spot position around element B. As a result, as shown in the inset of Fig. \ref{fig:DE}, the variation of the DE changed and now element B has the highest DE whereas element D shows the lowest DE. This means that the SSPD array actually has a spatial resolution because the photon-absorbed position can be identified from the peak-level variation in the output counts of the each element. The present time-resolved readout configuration should be suitable for various applications such as imaging sensors and spectroscopic systems, since this configuration can multiplex signals without losing position information. The total system DE of the present array system is 11.4\% since no crosstalk was observed. This value is comparable to that for practical single-pixel SSPD systems used for quantum key distribution demonstration \cite{Sasaki}. The DE will be improved further by adopting an optical-cavity structure and an interleaved layout of the array.

In conclusion, we have presented the successful operation of a four-element SSPD array integrated with a SFQ readout circuit in a compact GM cryocooler with a cooling capacity of 0.1 W. The time-resolved readout setup with a SFQ readout circuit clearly distinguished each element signal. We have verified that there was no crosstalk between elements even when connected with the SFQ circuit. This readout configuration also can be applied to imaging sensors and so on with spatially-addressed SSPD arrays. Our results formed the basis for the development of practical multi-element SSPD array systems. The improvement and evaluation of total system performance will be considered in a future work.

\end{document}